# Scanning tunneling microscopy and spectroscopy of nanoscale twisted bilayer graphene

Wen-Xiao Wang[1], Hua Jiang[2], Yu Zhang[1], Si-Yu Li[1], Haiwen Liu[1], Xinqi Li[3], Xiaosong Wu[3], and Lin He[1,*]

[1]Center for Advanced Quantum Studies, Department of Physics, Beijing Normal University, Beijing, 100875, People's Republic of China

[2]College of Physics, Optoelectronics and Energy, Soochow University, Suzhou, 215006, People's Republic of China

[3]State Key Laboratory for Artificial Microsctructure and Mesoscopic Physics, Peking University, Beijing 100871, China and Collaborative Innovation Center of Quantum Matter, Beijing 100871, China

* Email: helin@bnu.edu.cn

**Nanoscale twisted bilayer graphene (TBG) is quite instable and will change its structure to Bernal (or AB-stacking) bilayer with a much lower energy. Therefore, the lack of nanoscale TBG makes its electronic properties not accessible in experiment up to now. In this work, a special confined TBG is obtained in the overlaid area of two continuous misoriented graphene sheets. The width of the confined region of the TBG changes gradually from about 22 nm to 0 nm. By using scanning tunnelling microscopy, we studied carefully the structure and the electronic properties of the nanoscale TBG. Our results indicate that the low-energy electronic properties, including twist-induced van Hove singularities (VHSs) and spatial modulation of local density-of-state, are strongly affected by the translational symmetry breaking of the nanoscale TBG. Whereas, the electronic properties above the energy of the VHSs are almost not influenced by the quantum confinement even when the width of the TBG is reduced to only a single moiré spot.**

Nanoscale materials usually exhibit quite different properties from their bulk phase. For example, nanosized graphene structures, such as graphene nanoribbon and graphene quantum dot [1-10], show distinct electronic and magnetic properties from a continuous graphene monolayer. However, some of the materials do not allow to reduce their sizes to nanoscale due to the instability with reducing the size. Twisted bilayer graphene (TBG), composing of two mutually misoriented graphene layers with a rotation angle $\theta$ [11-25], is a typical example. Recent experiments demonstrated that nanosized TBG is not stable and will change its structure to nanosized Bernal (or AB-stacking) bilayer, which has a much lower energy [26-29]. Therefore, although the TBG have been studied extensively during the past few years [11-25], the electronic properties of the nanoscale TBG have remained experimentally elusive up to now.

In this work, a special nanosized TBG is obtained in the overlaid area of two continuous misoriented graphene sheets. The width of the confined region of the TBG, i.e., the overlaid area, changes gradually from about 22 nm to 0 nm. Such a structure is quite stable and provides us unprecedented opportunities to systematically study the electronic properties of the nanoscale TBG.

In our experiment, the graphene bilayer was synthesized on the SiC(000-1) by thermal decomposition (see Supplemental Material [30] for details). Previous studies indicated that adjacent graphene layers in such a system tend to have a strong twisting [31-33]. The characterizations and electronic properties of the sample were measured by using a ultra-high vacuum scanning tunneling microscopy (STM, Unisoku) (see Supplemental Material [30] for method and more information). Figure 1a shows a representative STM image of a TBG on the SiC(000-1). The period of the moiré pattern $D$ is about 4.9 nm and the twisted angle is estimated to be $\theta \approx 2.9º$ according to $D = a/[2\sin(\theta/2)]$ with $a = 0.246$ nm the lattice constant of graphene. The twist in real space also leads to the relative shift of the Dirac points (inset of Fig. 1(b)) on the different layers in the momentum space $\Delta K = 2|K|\sin(\theta/2)$, where $K$ is the reciprocal-lattice vector. Because of the interlayer coupling between the two adjacent graphene sheets, two saddle points appear at the intersections of the two Dirac cones, which consequently generate two low-energy van Hove singularities (VHSs) in the density-of-state (DOS). Figure 1(b) shows typical scanning tunneling spectroscopy (STS) spectra of the TBG. The two peaks with an energy spacing of about 430 meV in the spectra are attributed to the two VHSs, which agree well with both our theoretical result in Fig. 1 (c) and that reported in previous experimental work with a similar twisted angle [13-16].

The existence of the twist-induced moiré pattern is expected to affect the spatial distribution of local DOS (LDOS) in the TBG [13,16,24,25]. To further explore the effects of the moiré pattern on the electronic properties of the TBG, we carried out STS maps (differential conductance maps) measurements at various energies, which directly reflect the spatial distribution of the LDOS. Figure 2(a) (upper panel) shows a representative STS map recorded at energy of one of the VHSs of the TBG with $\theta \approx 2.9^\circ$ (see Fig. S1 [30] for more experimental data). Obviously, the LDOS in the TBG is strongly modulated by the moiré pattern and it reveals the same period of the moiré pattern, reflecting the presence of the moiré potential in the TBG. Figure 2(b) (upper panel) shows a STS map measured at an energy larger than that of the VHSs. The obtained LDOS also exhibits the same period of the moiré pattern, however, the positions of the strongest STS signal shift $D$/2, as shown in Fig. 2(c). In Fig. 2(a), the signal of the LDOS is strongest in the AA regions (where the sublattices of the adjacent graphene bilayer are aligned), whereas, the STS signal in Fig. 2(b) is strongest in the AB/BA regions (where Bernal stacking of the top and bottom graphene sheets occurs). Our results in Fig. 2 indicate that the wave functions of electrons in the TBG may switch the positions of their nodes at different energies. Such a behaviour may be related to the generation of moiré band in the TBG [20]. To fully understand the observed phenomena, we calculated the spatial distributions of the LDOS at the energies that we carried out measurements of the STS maps (see Supplemental Material [30] for details). The lower panels of Fig. 2(a) and 2(b) show the corresponding simulated results, which reproduce the main features of our experimental results quite well.

The above results show two important electronic properties, including the low-energy VHSs and the spatial modulation of the LDOS associated with the moiré pattern, of the TBG. We will show effects of quantum confinement on the two electronic properties of the TBG subsequently. To overcome the instability of the nanoscale TBG, here the nanoscale TBG is obtained in the overlaid area of two continuous graphene sheets with a small twisted angle, as shown in Fig. 3(a) (see Fig. S2 of Supplemental Material [30] for more STM measurements of the structure). Figure 3(b) and 3(c) show schematic diagrams of the unique nanoscale TBG. The period of the moiré pattern is about 6.6 nm and the twisted angle is estimated to be about 2.1°. As shown in Fig. 3(a), the width of the nanoscale TBG decreases from about 22 nm to 0 nm. Such a unique confined structure and, very importantly, its stability enable us to systematically study the electronic properties of the nanoscale TBG.

To further study the effect of quantum confinement on the electronic properties of the nanoscale TBG, we measured the tunneling spectra of the TBG with different width, as shown in Fig. 3(d). Obviously, the spectra depend sensitively on both the positions in the moiré pattern and the width of the nanoscale TBG. For the spectrum recorded at the position of the TBG with width of about 22 nm, two pronounced maxima of the LDOS at about -45 mV and 70 mV, corresponding to the two VHSs (also see Fig. S3 [30]), are clearly observed. According to the result in Fig. 3(d), the intensities of the two VHSs on the one hand are modulated periodically by the moiré pattern, and on the other hand decrease gradually with decreasing the width of the TBG (see Fig. S4 [30] for more experimental data). When the width of the TBG becomes comparable to that of a moiré spot (~ 6.6 nm), the intensities of the two VHSs become too weak to be observed. Such a result is reasonable because that the translational symmetry of the twist-induced moiré pattern is entirely removed in a single moiré spot. Consequently, the band structure based on the periodic moiré pattern should be completely destroyed. This behavior also can be understood within the framework of quantum confinement. For massless Dirac fermions confined in a nanoscale structure with the width $L$, the average level spacing of electronic states generated by quantum confinement can be estimated by $\Delta E \sim \alpha \pi \hbar v_F / L$, where $\alpha$ is a dimensionless constant of order unity, $\hbar$ is the reduced Planck's constant, and $v_F = 1.0 \times 10^6$ m/s is the Fermi velocity [34]. In our experiment, we observed that full width at half maximum (FWHM) of the VHSs increases with decreasing the width of the nanoscale TBG. For $L \sim 6.6$ nm, $\Delta E$ is estimated to be about 150 meV, which is sufficient large to broaden the VHSs to be invisible in the spectrum, in good agreement with our experimental result.

The effects of quantum confinement on the electronic properties of the TBG can be directly imaged by STS maps, as shown in Fig. 4. Figure 4(b) shows spatial distributions of the LDOS recorded at two different energies, one at − 45 meV (the energy of a VHS) and the other at 245 meV (the energy much larger than the VHS), along the nanoscale TBG. Two important results can be obtained according to the two STS maps shown in Fig. 4(b). First, the LDOS in the nanoscale TBG with width larger than 20 nm ($L \geq 3D$) is also strongly modulated by the moiré pattern and shows the same period as the moiré pattern (Fig. 4(a)). Such a behaviour is almost the same as that observed in a continuous TBG, as we have shown in Fig. 2. Second, the spatial distribution of the LDOS measured at − 45 meV (the energy of a VHS) is strongly disturbed by the quantum confinement when the width of the TBG is smaller than 20 nm. The period of the LDOS almost disappears when the width of the TBG is smaller than 10 nm (Fig. 4(c)). This behavior is similar as the width dependence of the intensities (or FWHM) of the VHSs, as we have shown in Fig. 3(d). Whereas, the spatial distribution of the LDOS recorded at 245 meV (above the energies of the VHSs) is only slightly influenced by the quantum confinement. The period of the LDOS maintains even when the width of the TBG is reduced to only a single moiré spot, as shown in Fig. 4(c). Such a result indicates that the observed spatial distributions of the LDOS at the two different energies have quite different origins. The spatial distribution of the LDOS at the VHSs should be strongly related to the band structure of the TBG. Therefore, it is sensitive to the translational symmetry breaking of the twist-induced moiré patterns. The moiré pattern will generate the moiré potential in the TBG, which can affect the spatial distribution of the LDOS. The observed behavior of the spatial distribution of the LDOS at high energies suggests that the spatial distribution of the LDOS at high energies may mainly originate from the moiré potential in the TBG and is weakly related to the band structure of the TBG.

In summary, a special nanosized TBG is obtained in the overlaid area of two continuous misoriented graphene sheets and we studied carefully the effects of quantum confinement on the electronic properties of the nanoscale TBG. Our results indicate that the VHSs and the spatial modulation of the low-energy LDOS are strongly affected by the translational symmetry breaking of the nanoscale TBG. Whereas, the electronic properties above the energy of the VHSs are almost not influenced by the quantum confinement even when the width of the TBG is reduced to only a single moiré spot.

# Reference

[1] K. Nakada, M. Fujita, G. Dresselhaus, M. S. Dresselhaus, Edge state in graphene ribbons: Nanometre size effect and edge shape dependence. *Phys. Rev. B* **54**, 17954 (1996).

[2] Y.-W. Son, M. L. Cohen, S. G. Louie, Half-metallic graphene nanoribbons. *Nature* **444**, 347 (2006).

[3] L. Yang, C. H. Park, Y. W. Son, M. L. Cohen and S. G. Louie, Quasiparticle energies and band gaps in graphene nanoribbons. *Phys. Rev. Lett.* **99**, 186801 (2007).

[4] G. Z. Magda, X. Jin, I. Hagymasi, P. Vancso, Z. Osvath, P. Nemes-Incze, C. Hwang, L. P. Biro and L. Tapaszto, Room-temperature magnetic order on zigzag edges of narrow graphene nanoribbons. *Nature* **514**, 608 (2014).

[5] Y. Y. Li, M. X. Chen, M. Weinert and L. Li, Direct experimental determination of onset of electron-electron interactions in gap opening of zigzag graphene nanoribbons. *Nature Commun.* **5**, 4311 (2014).

[6] W.-X. Wang, M. Zhou, X. Li, S.-Y. Li, X. Wu, W. Duan and L. He, Energy gaps of atomically precise armchair graphene sidewall nanoribbons. *Phys. Rev. B* **93**, 241403(R) (2016).

[7] S.-Y. Li, M. Zhou, J.-B. Qiao, W. Duan and L. He, Wide-band-gap wrinkled nanoribbon-like structures in a continuous metallic graphene sheet. *Phys. Rev. B* **94** (2016).

[8] S. K. Hamalainen, Z. Sun, M. P. Boneschanscher, A. Uppstu, M. Ijas, A. Harju, D. Vanmaekelbergh and P. Liljeroth, Quantum-confined electronic states in atomically

well-defined graphene nanostructures. *Phys. Rev. Lett.* **107**, 236803 (2011).

[9] D. Subramaniam, *et al*. Wave-function mapping of graphene quantum dots with soft confinement. *Phys. Rev. Lett.* **108**, 046801 (2012).

[10] S. K. Hamalainen, *et al*. Quantum-confined electronic states in atomically well-defined graphene nanostructures. *Phys. Rev. Lett.* **107**, 236803 (2011).

[11] J. M. B. Lopes dos Santos, N. M. R. Peres, A. H. Castro Neto, Graphene bilayer with a twist: electronic structure. *Phys. Rev. Lett.* **99**, 256802 (2007).

[12] S. Shallcross, S. Sharma, O. A. Pankratov, Quantum Interference at the Twist Boundary in Graphene. *Phys. Rev. Lett.* **101**, 056803 (2008).

[13] G. Li, A. Luican, J. M. B. Lopes dos Santos, A. H. Castro Neto, A. Reina, J. Kong and E. Y. Andrei, Observation of Van Hove singularities in twisted graphene layers. *Nat. Phys.* **6**, 109 (2010).

[14] W. Yan, M. Liu, R.-F. Dou, L. Meng, L. Feng, Z.-D. Chu, Y. Zhang, Z. Liu, J.-C. Nie, L. He, Angle-Dependent van Hove Singularities in a Slightly Twisted Graphene Bilayer. *Phys. Rev. Lett*. **109**, 126801 (2012).

[15] I. Brihuega, P. Mallet, H. Gonzalez-Herrero, G. Trambly de Laissardiere, M. M. Ugeda, L. Magaud, J. M. Gomez-Rodriguez, F. Yndurain and J. Y. Veuillen, Unraveling the intrinsic and robust nature of van Hove singularities in twisted bilayer graphene by scanning tunneling microscopy and theoretical analysis. *Phys. Rev. Lett.* **109**, 196802 (2012).

[16] L.-J. Yin, J.-B. Qiao, W.-J. Zuo, W.-T. Li, L. He, Experimental evidence for non-Abelian gauge potentials in twisted graphene bilayers. *Phys. Rev. B* **92**, 081406(R) (2015).

[17] E. Suarez Morell, J. D. Correa, P. Vargas, M. Pacheco, Z. Barticevic, Flat bands in slightly twisted bilayer graphene: tight-binding calculations. *Phys. Rev. B* **82**, 121407(R) (2010).

[18] T. Ohta, J. T. Robinson, P. J. Feibelamn, A. Bostwick, E. Rotenberg, T. E. Beechem, Evidence for Interlayer Coupling and Moiré Periodic Potentials in Twisted Bilayer Graphene. *Phys. Rev. Lett*. **109**, 186807 (2012).

[19] W.-Y. He, Z.-D. Chu, & L. He, Chiral Tunneling in a Twisted Graphene Bilayer. *Phys. Rev. Lett.* **111**, 066803 (2013).

[20] R. Bistritzer, & A. H. MacDonald, Moire bands in twisted double-layer graphene. *Proc Natl Acad Sci* (*USA*) **108**, 12233-12237 (2011).

[21] P. San-Jose, J. Gonzalez, F. Guinea, Non-Abelian gauge potentials in graphene bilayers. *Phys. Rev. Lett.* **108**, 216802 (2012).

[22] W. Yan, L. Meng, M. Liu, J.-B. Qiao, Z.-D. Chu, R.-F. Dou, Z. Liu, J.-C. Nie, D. G. Naugle, L. He, Angle-dependent van Hove singularities and their breakdown in twisted graphene bilayers. *Phys. Rev. B* **90**, 115402 (2014).

[23] Z.-D. Chu, W.-Y. He, L. He, Coexistence of van Hove singularities and superlattice Dirac points in a slightly twisted graphene bilayer. *Phys. Rev. B* **87**, 155419 (2013)

[24] D. Wong, Y. Wang, J. Jung, S. Pezzini, A. M. DaSilva, H. Tsai, H. S. Jung, R. Khajeh, Y. Kim, J. Lee, S. Kahn, S. Tollabimazraehno, H. Rasool, K. Watanabe, T. Taniguchi, A. Zettl, S. Adam, A. H. MacDonald, M. F. Crommie, Local spectroscopy of moiré-induced electronic structure in gate-tunable twisted bilayer graphene. *Phys. Rev. B* **92**, 155409 (2015).

[25] L.-J. Yin, J.-B. Qiao, W.-X. Wang, W.-J. Zuo, W. Yan, R. Xu, R.-F. Dou, J.-C. Nie and L. He, Landau quantization and Fermi velocity renormalization in twisted graphene bilayers. Phys. Rev. B **92**, 201408 (2015).

[26] M. Zhu *et al.*, Stacking transition in bilayer graphene caused by thermally activated rotation. *2D Materials* **4**, 011013 (2016).

[27] D. Wang *et al.*, Thermally Induced Graphene Rotation on Hexagonal Boron

Nitride. *Phys. Rev. Lett.* **116**, 126101 (2016).

[28] E. Koren, I. Leven, E. Lortscher, A. Knoll, O. Hod and U. Duerig, Coherent commensurate electronic states at the interface between misoriented graphene layers. *Nature Nano.* **11**, 752 (2016).

[29] C. R. Woods *et al.*, Macroscopic self-reorientation of interacting two-dimensional crystals. *Nature Commun.* **7**, 10800 (2016).

[30] See Supplemental Material for methods, more STM images, STS spectra, and details of the calculations.

[31] T. Cai, Z. Jia, B. Yan, D. Yu and X. Wu, Hydrogen assisted growth of high quality epitaxial graphene on the C-face of 4H-SiC. *App. Phys. Lett.* **106**, 013106 (2015).

[32] J. Hass *et al.*, Why Multilayer Graphene on 4H−SiC(000-1) Behaves Like a Single Sheet of Graphene. *Phys. Rev. Lett.* **100**, 125504 (2008).

[33] F. Varchon, P. Mallet, L. Magaud and J.-Y. Veuillen, Rotational disorder in few-layer graphene films on 6H−SiC(000−1): A scanning tunneling microscopy study. *Phys. Rev. B* **77**, 165415 (2008).

[34] Y. Zhao *et al.*, Creating and probing electron whispering-gallery modes in graphene. Science **348**, 672 (2015).

**Acknowledgments**

This work was supported by the National Natural Science Foundation of China (Grant Nos. 11674029, 11422430, 11374035, 11374219, 11504008), the National Basic Research Program of China (Grants Nos. 2014CB920903, 2013CBA01603, 2014CB920901), the NSF of Jiangsu province, China (Grant No. BK20160007), the program for New Century Excellent Talents in University of the Ministry of Education of China (Grant No. NCET-13-0054). L.H. also acknowledges support from the National Program for Support of Top-notch Young Professionals and support from "the Fundamental Research Funds for the Central Universities".

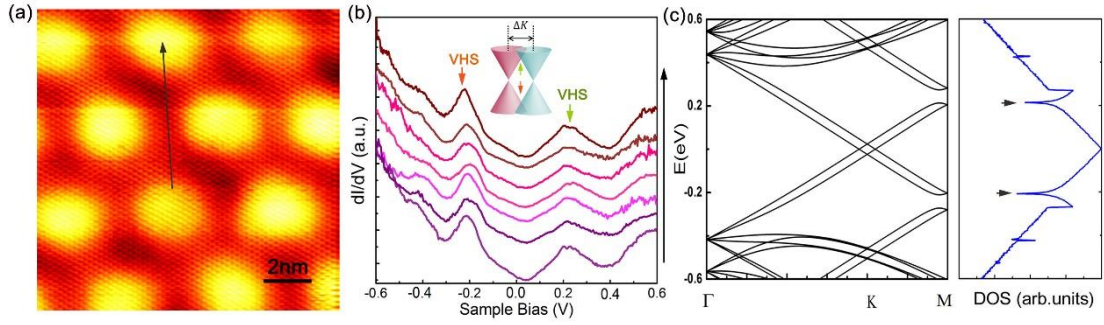

**FIG. 1.** (a) A typical STM topography of graphene bilayer with twisted angle $\theta \sim 2.9°$ ($V_{sample}$ = 223 mV and I = 180 pA). The period of moiré pattern is about 4.9 nm. (b) Tunneling spectra recorded along the arrow of (a). (c) Theoretical calculated band structure (left panel) of the twist bilayer graphene with the twisted angle $\theta \sim 2.9°$ and the corresponding LDOS (right panel) with two VHSs (pointed out by the arrows).

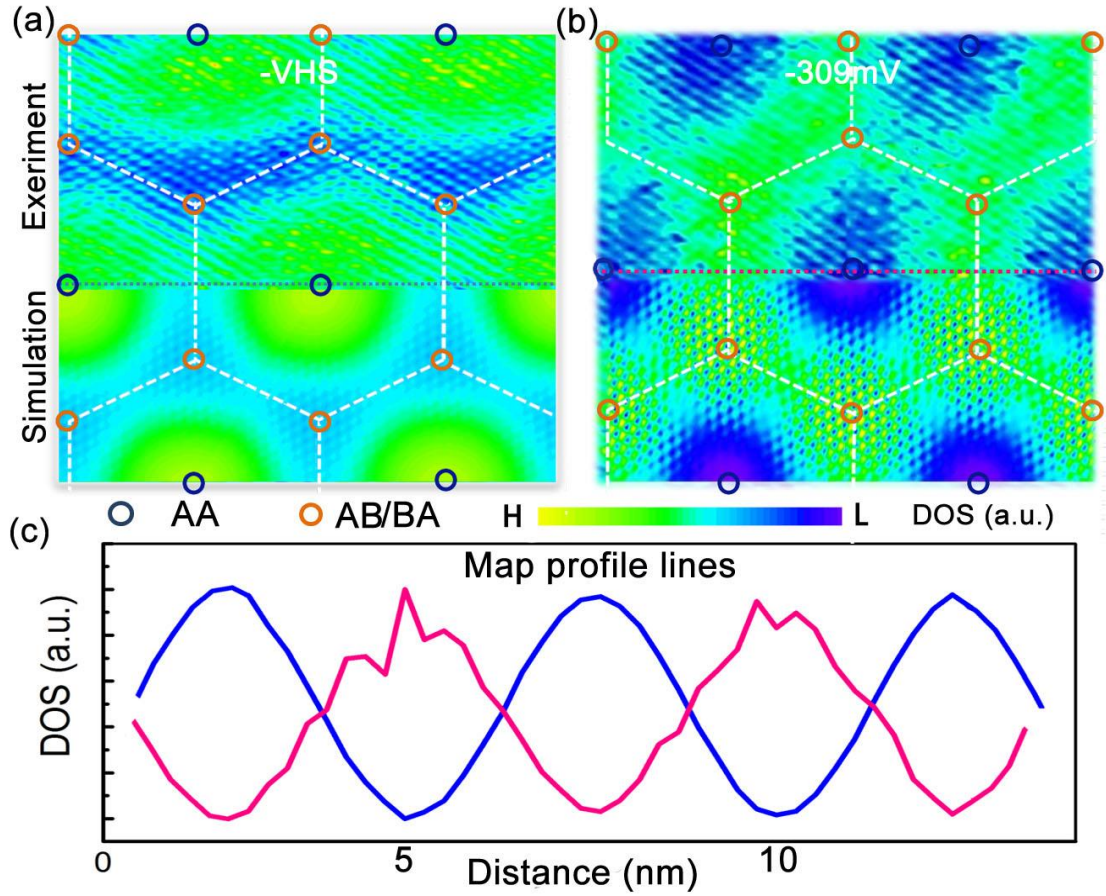

**FIG. 2.** (a), (b) Conductance maps (upper panels) of the TBG with the twisted angle $\theta$ ~ 2.9 ° measured at -VHS ($V_{sample}$ = -210 mV) and -309 mV, respectively. Lower panels of (a) and (b) are the corresponding theoretical results. (c) The profile lines along the moiré pattern of the STS maps in (a) and (b). The positions of the strongest STS signal of (a) and (b) shift $D/2$.

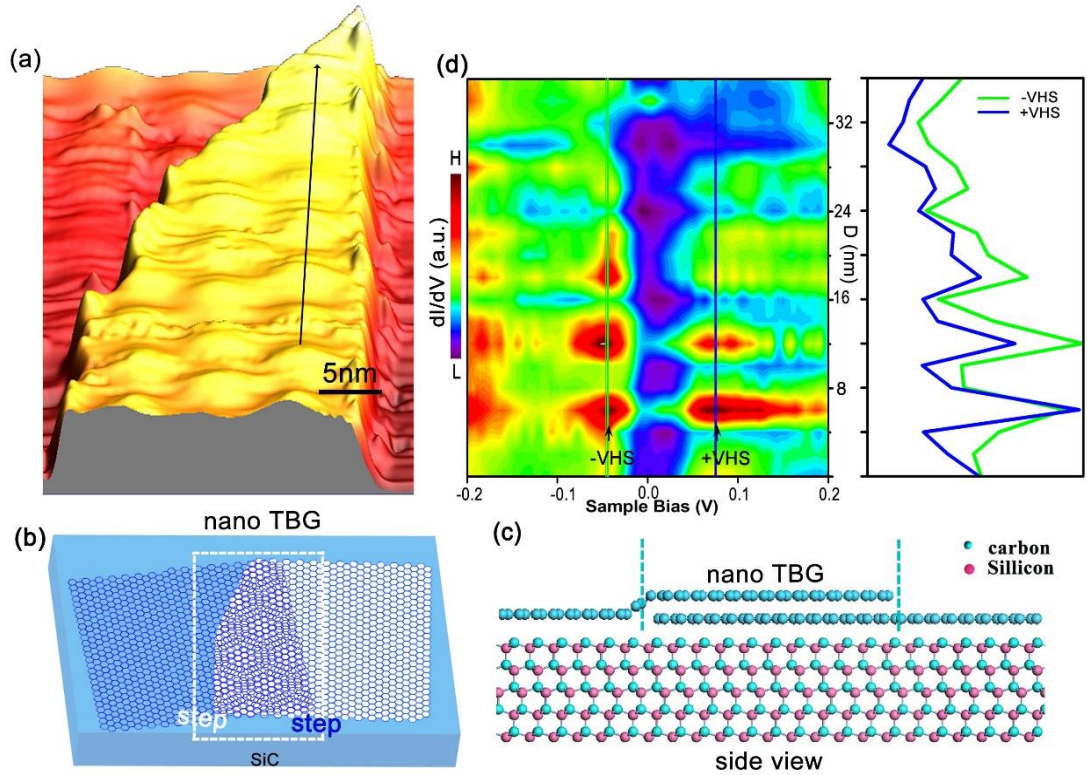

FIG.3 (a) STM image of a nanoscale TBG with width decreasing from 22 nm to 0 nm ($V_{sample}$ = -700 mV and I = 200 pA). The twist angle of the TBG is about 2.1° and the period of the moiré pattern is about 6.3 nm. (b) The schematic diagram to illustrate the structure of the confined TBG. (c) The side view of the nanoscale TBG on a SiC substrate. (d) The STS spectra measured along the arrow in (a). Right panel shows profile lines taken from the left panel. The intensities of the VHSs decreases gradually with decreasing the width of the TBG.

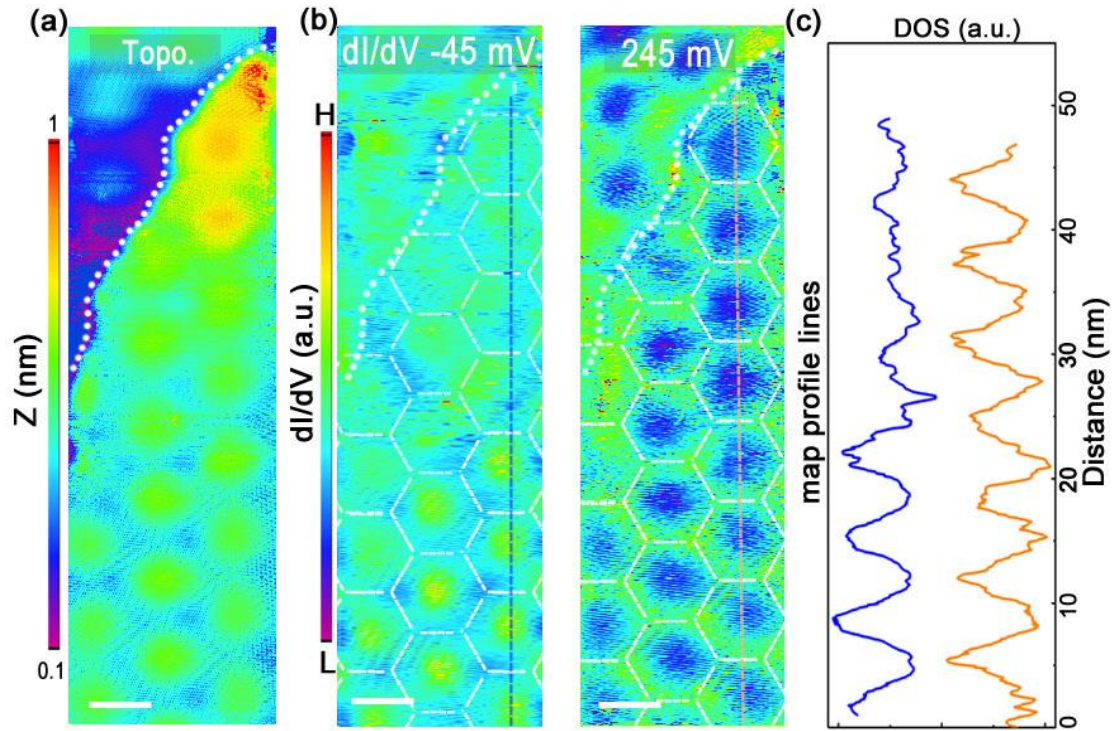

**FIG. 4.** (a) STM image of the nanoscale TBG with $\theta \sim 2.1°$ ($V_{sample}$ = 450 mV and I = 120 pA). (b) Conductance maps of the nanoscale TBG at energies - 45 meV and 245 meV. (d) The profile lines (blue and orange) taken from (b), explicitly showing the quantum confinement on the spatial distributions of the LDOS.